\documentclass{article}
\usepackage{hiph-art}
\volnumber{} \issuenumber{} \edyear{2005}                             
\frompage{000} \topage{000}                                              
\recrevdate{31 Oct 2005}                                              
\def\rightmark{ Strange quark collectivity of $\phi$ meson at RHIC}
\def\leftmark{J. H. Chen et al.}

\title{Strange quark collectivity of $\phi$ meson at RHIC}
\authors{{J. H. Chen $^{1,2}$, \underline{Y. G. Ma}$^{1,a}$, G. L. Ma $^{1,2}$,
 H. Z. Huang $^{1,3}$, X. Z. Cai$^1$, Z. J. He $^1$, J. L. Long $^1$,  W. Q. Shen
 $^1$, J. X. Zuo $^{1,2}$
}\\[2.812mm]
{\normalsize \hspace*{-8pt} $^1$  Shanghai Institute of Applied
Physics, Chinese Academy of
Sciences, Shanghai 201800, China \\[0.2ex] \hspace*{-10pt}
$^2$ Graduate School of the Chinese Academy of Sciences, Beijing
100080, China \\[0.2ex] \hspace*{-10pt}
$^3$ University of California, Los Angeles, CA90095, USA%
}}

\abstract{Based on A Multi-Phase Transport model, the elliptic
flow $v_{2}$ of $\phi$ mesons which is reconstructed from
$K^{+}K^{-}$ at the Relativistic Heavy Ion Collider (RHIC) energy
has been studied. The results show that reconstructed $v_{2}$ of
$\phi$ meson can keep the earlier information before $\phi$ decays
and it seems to obey the number of constituent quark scaling as
other mesons and baryons. This result indicates that the $\phi$
$v_2$ mostly reflects the parton level collectivity developed
during the early stage of the collisions and the strange and light
up/down quarks have similar collectivity properties before the
hadronization.}



\makeindex

\begin{document}

\maketitle

\section{Introduction}\label{intro}

Elliptic flow in heavy ion collisions is a measure of the
azimuthal anisotropy of particle momentum distribution in the
plane perpendicular to the beam direction \cite{Olli}. It results
from the initial spatial asymmetry in noncentral collision and the
subsequent collective interaction is thus sensitive to the
properties of the dense matter formed during the initial stage of
heavy ion collision \cite{Sorge,Danielewice,Kolb} 
and parton dynamics \cite{ZhangB} at RHIC energies.
 The experimental results
of charged kaons, protons and pions \cite{STAR1,PHENIX1} show that
the elliptic flow first increases with particle transverse
momentum following the hydrodynamic behavior and then tends to
saturation in intermediate transverse momentum region.  More
importantly, a Number-of-Constituent-Quark (NCQ) scaling
phenomenon  of the elliptic flow has been discovered from this
saturation region for baryons and mesons.

Strange quark dynamics is believed to be a useful probe of the QCD
matter created at RHIC. Enhanced strangeness production
\cite{Rafelski} has been proposed as an important signal for the
QGP phase transition. The dominant production of $s\overline{s}$
pairs via gluon-gluon interaction may lead to strangeness
(chemical and flavor) equilibration times comparable to the
lifetime of the plasma and much shorter than that of a thermally
equilibrated hadronic fireball. The subsequent hadronization is
then expected to result in an enhanced production of strangeness
particle. In particular, it has been argued that with the
formation of QGP not only the production of $\phi$-meson, which
consists of $s\overline{s}$, is enhanced but they also retain the
information on the condition of the hot plasma. It is believed
that $\phi$-meson interacts weakly in the hadronic matter and
therefore freezes out quite early from the system \cite{Shor}.
Therefore the $\phi$-meson in RHIC has been of great interest
\cite{STAR,PHENIX}.

We use a multi-phase transport (AMPT)  model to investigate effect
of parton dynamics on $\phi$ meson. AMPT model is a hybrid model
which is consisted of four main components: the initial condition,
partonic interactions, the conversion from partonic matter into
hadronic matter and hadronic interactions. Details of the AMPT
model can be found in \cite{AMPT}. In the default AMPT model
\cite{DAMPT} partons are recombined with their parent strings when
they stop interaction, and the resulting strings are converted to
hadrons using a Lund string fragmentation model \cite{Lund}. In
the AMPT model with string melting \cite{SAMPT}, a simple quark
coalescence model is used to combine partons into hadrons.

In this work, we present a study for the elliptic flow of
$\phi$-meson at RHIC energy based on AMPT model with string
melting scenario. We illustrate  that the partonic effect could
not be neglected for $\phi$ mesons as demonstrated for other
hadrons in \cite{SAMPT} and a string melting AMPT version is much
more appropriate than the default AMPT version when the energy
density is much higher than the critical density for the pQCD
phase transition. So far, the AMPT model with string melting
scenario has been successful to describe the elliptic flow of
stable baryons and mesons \cite{AMPT,SAMPT}. In this work, we try
to investigate an unstable particle: $\phi$ meson. In order to
compare with the experimental data, we adopt the value of particle
mass with the mass width according to Breit-Wigner shape and then
a broadening width of Breit-Wigner shape has been obtained when we
reconstruct $\phi$ \cite{PYTHIA}. The value of the parton
scattering cross section is chosen as 10 mb. The transverse
momentum dependence and the collision centrality dependence  have
been studied and the NCQ-scaling phenomenon has been observed for
$\phi$-mesons. In addition, the rescattering effect of $\phi$ flow
has been investigated in the hadronic scattering model, namely ART
model which has been modified to include $\phi$ meson scattering
processes \cite{Pal}.

\section{Analysis Method}\label{analysis}

The azimuthal distribution with respect to the reaction plane at
rapidity $y$ can be written in a form of Fourier series as follows
\begin{equation}
E\frac{d^{3}N}{d^{3}p}=
\frac{1}{2\pi}\frac{d^{2}N}{p_{T}dp_{T}dy}(1+\sum_{n=1}^{\infty}2v_{n}cos[n(\Phi-\Psi)]),
\end{equation}
where $p_T$ is transverse momentum,  $\Phi$ is azimuthal angle and
$\Psi$ is the reaction plane angle. The first and second Fourier
coefficients $v_{1}$ and $v_{2}$ are called directed and elliptic
flow, respectively.

\begin{figure}[htb]
\vspace*{-1.1cm}
                 \insertplot{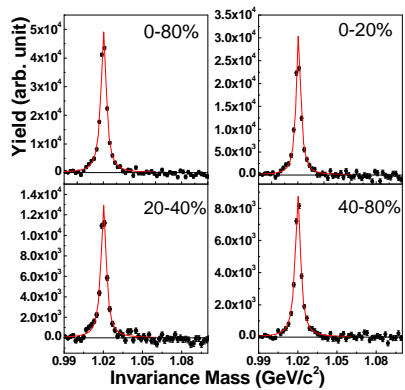}
\vspace*{-1.1cm} \caption[]{The reconstructed invariant mass
distribution (dots with statistical error bar) with the fits of
Breit-Wigner function plus a linear background in the mass
(GeV/$c^{2}$) range of (0.98, 1.06) of $\phi$-meson from
$K^{+}K^{-}$ pairs. From top-left to bottom-right it corresponds
to the centrality of 0-80$\%$, 0-20$\%$, 20-40$\%$ and 40-80$\%$,
respectively. The azimuth angle range is 0-$\pi$ and the $p_T$
(GeV/c) range is 0.4 - 2.4 GeV/c. The FWHM from fitted function is
5.003, 5.073, 5.004 and 4.869 MeV respectively while the default
value in AMPT model is 4.43 MeV.}
 \label{fig_inv}
\end{figure}

Since $\phi$-meson is unstable, it can only be reconstructed in
final state from its decay products of either the $K^{+}K^{-}$
pair or the lepton ($e^{+}e^{-}$ or $\mu^{+}\mu^{-}$) pair. The
present simulation results are reconstructed from the former decay
products among $\sim$ 100,000 events. All identified $K^{+}$ and
$K^{-}$ particles in a given event within the rapidity range of
(-1,1) are combined to form the invariant mass distribution. There
are a large number of combinatorial background when the invariant
mass distribution is reconstructed by $K^{+}K^{-}$ pair. The
combinatorial background are estimated by an event mixing method
\cite{MIX} in which all $K^{+}$ particles from one event are
combined with $K^{-}$ particles of ten other events within the
same centrality. Then, the number of the combinational background
are obtained after the normalized factor of the mixed event number
(ten in our analysis). Finally, the yield in each bin is
determined by fitting the background subtracted invariant mass
distribution to a Breit-Wigner function plus a linear background
in a limited mass range. Fig.~\ref{fig_inv} shows some examples in
certain transverse momentum ranges for the reconstructed invariant
mass spectra of $\phi$ mesons.

The magnitude of the anisotropy and the finite number of particles
available to determine the reaction plane leads to a finite
resolution. Therefore, the reconstructed $v_{n}^{rec}$
coefficients with respect to the reaction plane have to be
corrected for the reaction plane resolution \cite{Danielewicz2}:
\begin{equation}
v_{n} = \frac{v_{n}^{res}}{\langle
cos(n(\Psi_{n}-\Psi_{R}))\rangle},
\end{equation}
where $\Psi_n$ is reconstructed event plane angle. The mean cosine
values are less than unity thus this correction always increases
the flow coefficients. After such correction, the flow should tend
to the "true" flow which is supposed to be determined in true
reaction plane.

\section{Results}\label{results}

\begin{figure}
\vspace*{-0.1cm}
                 \insertplot{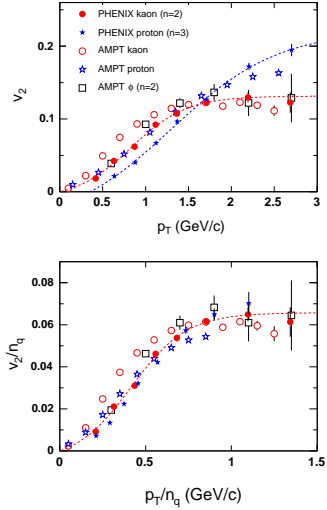}
\vspace*{-1.1cm} \caption[]{Top panel: $p_T$ dependence of the
$v_2$ for $\phi$-meson compared with $K^{+}$+$K^{-}$ and
$p+\overline{p}$. Data  are taken from Ref.~\cite{PHENIX1}.
Dot-dashed lines are the fits results of function ( $f_{v_{2}}$(n)
 = $\frac{a n}{1 + exp(-(p_T/n - b)/c)}$ - dn, where $a, b, c$ and $d$ are
the fit parameters, $n$ is the constituent-quark number ). Bottom
panel: NCQ scaled $v_{2}$. The error bars represent statistical
errors only.} \label{fig_ncq}
\end{figure}

The upper panel of Fig.\ref{fig_ncq} shows the elliptic flow
$v_{2}$ of $\phi$-meson from minimum-bias $^{197}$Au + $^{197}$Au
collisions at $\sqrt{s_{NN}}$ = 200 GeV. Experimental data
\cite{PHENIX1} of $K^{+}$+$K^{-}$ and $p+\overline{p}$ are also
presented for comparison. First we observe that  $v_{2}$ of
$\phi$-meson seems to follow the same behavior of $K^{+}$+$K^{-}$:
it is higher than baryon at low $p_T$ region and then saturate in
intermediate $p_T$ region. The $\phi$-meson $v_{2}$ is in
agreement with the hydrodynamic model calculation \cite{Huovinen}
which predicts its mass ordering at low $p_T$ region - perhaps
implying that an early thermalized system has been created in
collisions at RHIC energy. Compared with experimental data of
$K^{+}$+$K^{-}$ and $p+\overline{p}$, the reconstructed result can
give a good description of the $\phi$-meson $v_{2}$ within error
bars.

One of the salient observations made at RHIC is a NCQ-scaling of
the hadronic elliptic flow at intermediate $p_T$ (1.2 $<p_T< $ 4.0
GeV/c) \cite{STAR1} and  the quark coalescence or recombination
mechanism has been used to explain the NCQ-scaling
\cite{Lin2,Fries} in that $p_T$ region. In this coalescence or
recombination mechanism, hadron is formed via  quark coalescence.
The lower panel of Fig.~\ref{fig_ncq} displays angular anisotropy
of the constituent quark based on  the quark coalescence picture.
The behavior is similar for all reference particles for
$p_T$/$n_{q}$ $>$ 0.6 GeV/c. The $v_{2}$($p_T$/$n_{q}$) represents
a constituent quark momentum-space anisotropy $v_{2}^{q}$ that may
indicate a consequence of collectivity in a partonic stage which
includes strange quarks. The reconstructed results give an
implication that a new stage of partonic matter may be created
with $v_{2}^{q}$ characterizing the property of the matter.

As the $K^{+}+K^{-}$ pair decaying from a $\phi$-meson is likely
to undergo appreciable re-scattering in the medium and this might
lead to a reconstructed invariant mass situated outside the
original $\phi$-meson peak. In this case, the in-medium effect is
necessary to be studied in details. It is carried out through
turning off or turning on ART process \cite{ART} in AMPT model. It
was found that the in-medium hadronic re-scattering effect on the
final $\phi$ elliptic flow can be ignored within the errors. The
reason is that the spatial anisotropic during the hadronic phase
is small as well as $\phi$ has smaller hadronic interaction cross
section. This may, however, depend on the hadronization scheme
\cite{SAMPT}.  Our reconstructed result for $\phi$ $v_2$ confirms
that it can keep useful information of $\phi$-meson which is
produced during the earlier collision.

\section{Summary}\label{summary}

In summary, we have firstly presented the elliptic flow $v_{2}$ of
$\phi$ meson from minimum-bias $^{197}$Au+$^{197}$Au collisions at
$\sqrt{s_{NN}}$ = 200 GeV in a multi-phase transport  model with
string melting scenario. The $v_{2}$ of $\phi$-meson seems to
demonstrate the similar behavior to   other mesons. A NCQ-scaling
phenomenon of elliptic flow reveals for $\phi$ from the
reconstruction of $K^+ K^-$ pairs. The coefficient
$v_{2}$($p_T/n_{q}$) of $\phi$ represents a constituent strange
quark momentum-space anisotropy $v_{2}^{q}$ that may arise as a
consequence of collectivity in a partonic degree of freedom. It
may imply that a new state of partonic matter has been created
with $v_{2}^{q}$ characterizing the property of the matter. The
RHIC data for the elliptic flow of $\phi$ will shed light on these
physics issues.

\section*{Acknowledgments}
We would like to acknowledge Dr. C. M. Ko for discussions. This
work was partially supported by the Shanghai Development
Foundation for Science and Technology under Grant Numbers
05XD14021, the National Natural Science Foundation of China under
Grant No 10328259 and 10135030.

\section*{Notes}
\begin{notes}
\item[a] Speaker and Corresponding author.  E-mail:
ygma@sinap.ac.cn
\end{notes}

\vfill\eject
\end{document}